\documentclass[reprint,showpacs,superscriptaddress,preprintnumbers,aps,pra]{revtex4-1}
\usepackage{amsfonts, amsmath,mathptmx,amssymb,bbm, bbm, graphicx, epsfig, epstopdf, float, placeins}
\usepackage[dvipsnames]{xcolor}
\usepackage[colorlinks=true,linkcolor=Blue,citecolor=Blue,urlcolor=Blue,linktocpage=true]{hyperref}

\begin{document}
\title{Quantum correlations in optomechanical crystals}
\author{F. Bemani}  \email{foroudbemani@gmail.com}
\affiliation{Department of Physics, Faculty of Science, University of Isfahan, Hezar Jerib, 81746-73441, Isfahan, Iran}
\author{R. Roknizadeh} \email{r.roknizadeh@gmail.com}
\affiliation{Department of Physics, Faculty of Science, University of Isfahan, Hezar Jerib, 81746-73441, Isfahan, Iran}
\affiliation{Quantum Optics Group, Department of Physics, Faculty of Science, University of Isfahan, Hezar Jerib, 81746-73441, Isfahan, Iran}
\author{A. Motazedifard} \email{motazedifard.ali@gmail.com}
\affiliation{Department of Physics, Faculty of Science, University of Isfahan, Hezar Jerib, 81746-73441, Isfahan, Iran}
\author{M. H. Naderi} \email{mhnaderi@sci.ui.ac.ir}
\affiliation{Department of Physics, Faculty of Science, University of Isfahan, Hezar Jerib, 81746-73441, Isfahan, Iran}
\affiliation{Quantum Optics Group, Department of Physics, Faculty of Science, University of Isfahan, Hezar Jerib, 81746-73441, Isfahan, Iran}
\author{D. Vitali}\email{david.vitali@unicam.it}
\affiliation{Physics Division, School of Science and Technology, University of Camerino, I-62032 Camerino (MC), Italy}
\affiliation{INFN, Sezione di Perugia, via A. Pascoli, Perugia, Italy}
\affiliation{CNR-INO, L.go Enrico Fermi 6, I-50125 Firenze, Italy}
\date{\today}

\begin{abstract}
The field of optomechanics provides us with several examples of quantum photon-phonon interface. In this paper, we theoretically investigate the generation and manipulation of quantum correlations in a microfabricated optomechanical array. We consider a system consisting of localized photonic and phononic modes interacting locally via radiation pressure at each lattice site with the possibility of hopping of photons and phonons between neighboring sites. We show that such an interaction can correlate various modes of a driven coupled optomechanical array with well-chosen system parameters. Moreover, in the linearized regime of Gaussian fluctuations, the quantum correlations not only survive in the presence of thermal noise, but may also be generated thermally. We find that these optomechanical arrays provide a suitable platform for quantum simulation of various many-body systems.
\end{abstract}

\maketitle

\section{Introduction}
The impressive experimental progress in fabricating micromechanical and nanomechanical devices have opened a route towards the exhibition of quantum behavior at macroscopic scales. The interaction between micro- or nanomechanical oscillators and the optical field via the radiation pressure force is the basis of a wide variety of optomechanical phenomena. Despite their variety in the system sizes, parameters, and configurations, optomechanical systems (OMSs) share common features. Almost all OMSs are described by the same physics. OMSs offer further insights into the issues concerning the development of quantum memory for quantum computers \cite{Cole}, high precision position, mass or force sensing \cite{Krause,Motazedi,Murch,Purdy,Chaste}, quantum transducers \cite{Bochmann}, classical and quantum communication \cite{Palomaki}, ground state cooling of mechanical oscillators \cite{Chan2011,Teufel2011}, nonclassical correlations between single photons and phonons \cite{Riedinger}, generation of nonclassical states \cite{Brooks} and testing of the foundations of quantum mechanics \cite{Bawaj,Pikovski,Vivoli,Marinkovi}. For a recent review and current areas of focus of quantum optomechanics see Refs. \cite{Aspelmeyer,Bowen}.

The extension to multimode systems is an attractive route for quantum optomechanics. A group of mechanical oscillators interacting via the radiation pressure with a common optical mode \cite{Xuereb,Holmes,Xuereb2,Xuereb3,Seok,Bhattacharya,Tomadin,Bemani18}, or a group of mechanical oscillators locally interacting with a single optical mode involving the tunneling of photons and phonons between neighboring sites \cite{Heinrich,Ludwig,SafaviNaeini2011,Chan,Safavi,Chang2011,Ludwig2013,Chang2011,Schmidt,Chen,Schmidt2015,Schmidt15,Peano} are the two realizations of multimode optomechanics. The former is realized in a single optical cavity containing multiple membranes while the latter is realized experimentally in the so-called optomechanical crystals (OMCs) in one and two dimensions.

Cooperative behaviors, emerging due to the mutual coupling, are beneficial to investigate many-body physics of photons or phonons in OMCs. An OMC is usually fabricated from a thin film of silicon membranes where an engineered defect in the crystal is used to localize an optical and a mechanical mode. OMCs usually have a large single photon optomechanical coupling \cite{Eichenfield,SafaviNaeini2010,Gavartin,SafaviNaeini2014}. Several aspects of the array of coupled OMSs have already been investigated in the literature, involving synchronization dynamics \cite{Ludwig,Heinrich,Zhang,Bemani18,Mari}, slowing and stopping light \cite{Chang2011}, long-range collective interactions \cite{Xuereb}, correlated quantum many-body states \cite{Ludwig2013}, reservoir engineering and dynamical phase transitions \cite{Tomadin}, squeezing, entanglement and state transfer between modes \cite{Schmidt,Akram}, transport in a one-dimensional chain \cite{Chen,Gan,Xiong}, superradiance and collective gain \cite{Kipf}, graphene-like Dirac physics \cite{Schmidt2015}, creation of artificial magnetic fields for photons on a lattice \cite{Schmidt15}, quantum simulation of the propagation of the collective excitations of the photon fluid in a curved spacetime \cite{Bemani:17}, and topological phases of sound and light \cite{Peano}.

Quantum correlations, in particular entanglement, have many applications in superdense coding, quantum teleportation \cite{Bouwmeester} and protocols of quantum cryptography \cite{Ekert1991}. The generation and manipulation of entanglement in many-body systems are of great importance for quantum information processing. Furthermore, quantum correlations are valuable in characterizing various phases and corresponding quantum phase transitions in quantum many-body systems \cite{Dillenschneider,Sylju,Osborne}. Bipartite entanglement plays an important role in characterizing, classifying and simulating quantum many-body systems \cite{Chiara}. Physical systems such as Bose-Einstein condensates \cite{Sorensen,Esteve,Eckert2008}, cold or thermal atoms \cite{Chaudhury,Fernholz}, and trapped ions \cite{Leibfried,Molmer} represent promising platforms for the investigation of many-particle quantum entanglement. In the past decade, much of the attention has been devoted to entanglement in OMSs. Entanglement is one of the consequences of the coherent photon-phonon interaction in OMSs \cite{Muller,Hartmann,Vitali,Vitali1,Palomaki,Paternostro,Mazzola,Barzanjeh}. For instance, continuous variables entanglement between two mechanical modes has recently been realized \cite{Riedinger2,Ockeloen-Korppi2018}.  Since it is a possible resource for quantum technologies, quantum discord in many-body systems also requires attention.

Despite considerable efforts to understand the quantum correlations in OMSs \cite{Muller,Hartmann,Vitali,Vitali1,Palomaki,Paternostro,Mazzola,Riedinger2,Ockeloen-Korppi2018}, a full picture of the behavior of entanglement and of quantum discord in OMCs remain elusive. Based on the above motivations, in this paper, we consider the dynamics of coupled OMSs with a view towards quantum correlations. Employing the Heisenberg-Langevin (HL) approach and linearizing HL equations, we separate the deterministic dynamics and the quantum fluctuation dynamics. We then use HL equations to obtain the covariance matrix (CM) in order to study quantum correlations. With the CM in hand, we can investigate the degree of steady-state entanglement and the Gaussian quantum discord between different optical and mechanical modes under different conditions. We study the influence of the presence of a thermal reservoir and we show a nonmonotonic behavior of quantum correlations as a function of the heat bath temperature.

The paper is organized as follows. In Sec. \ref{sec:Sec2}, we begin with describing the system under consideration, i.e., an OMC. In Sec. \ref{sec:Sec3}, we derive the HL equations of motion.  We then discuss the classical equations of motion and the linearized quantum equations. In Sec. \ref{sec:Sec4}, we discuss the presence of entanglement and Gaussian discord in OMCs. Finally, in Sec. \ref{sec:Sec5}, we present our concluding remarks.
\section{\label{sec:Sec2} Array of coupled OMS$\textrm{s}$}
\begin{figure}
	\includegraphics[scale=1]{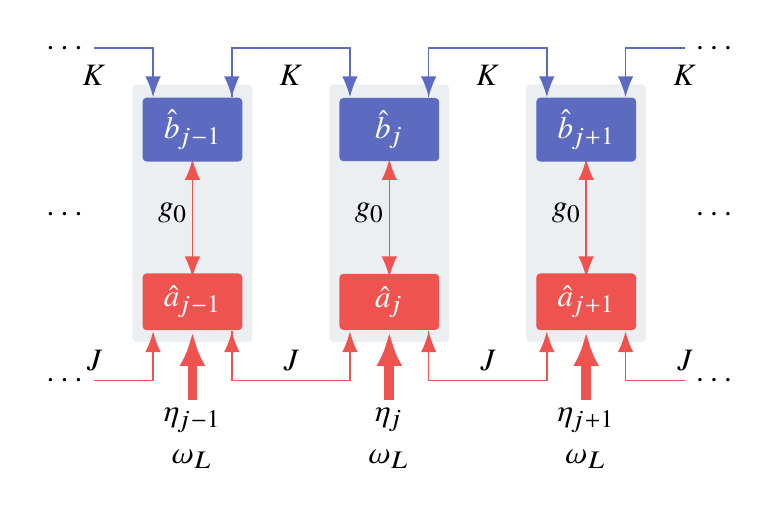}
	\caption{Schematic illustration of  a one-dimensional OMC. Localized photonic and phononic modes couple locally via the optomechanical interaction with strength $ g_0$ at each lattice site.  Photon and phonons hop between near neighbor sites with rates $J$ and $K$, respectively. A laser at frequency $ \omega_L $ and amplitude $ \eta_j $ drives  each site.}
	\label{Fig:Fig1}
\end{figure}
As depicted in Fig.~\ref{Fig:Fig1}, the system under consideration is a finite one-dimensional OMC where each site consists of a localized photonic and phononic mode coupled locally via the standard optomechanical interaction. The modes of nearby sites are connected via photon and phonon tunneling. The Hamiltonian of such a system is then given by ($ \hbar=1 $) \cite{Ludwig,Chen,Peano,Gan}
\begin{equation}
{H}  = H_0+ H_t+ H_p\,,
\label{Eq:Optomechanical_Array_Hamiltonian}
\end{equation}
where
\begin{subequations}
	\begin{align}
	H_0 &=\sum\limits_j {\left[ {{\omega _c} a_j^\dag {a_j} + {\omega _m} b_j^\dag {b_j} - {g_0} a_j^\dag {a_j}( b_j^\dag  + {b_j})} \right]},\\
	H_t&=- \sum\limits_{\left\langle {j,l} \right\rangle } \big({{J} a_j^\dag } {a_l} +{K b_j^\dag } {b_l}\big)\,,\\
	H_p&= \sum\limits_j {\big({\rm i}\eta _je^{ - {\rm i}\omega _L t} a_j^\dag  - {\rm i}\eta _j^*e^{{\rm i}\omega _Lt}  a_j\big)} \,.
	\end{align}
\end{subequations}
Here, $  H_0 $ includes the free energy of each optical mode with frequency $\omega_c$, denoted by the photon operators ${a}_{j}$ and ${a}_{j}^\dag$, the harmonic motion of each mechanical modes with frequency $\omega_m$, denoted by phonon operators ${b}_{j}$ and ${b}_{j}^\dag$, and the usual optomechanical interaction with strength $g_0$.
Further, $ {H}_t $ represents the hopping of photons and phonons between adjacent lattice sites with hopping strengths $J$ and $K$, respectively. The notation $\sum\nolimits_{\left\langle {j,l} \right\rangle } {} $ denotes the summation over all adjacent lattice sites.
Finally, $ H_p$ denotes that each lattice site is optically driven by a laser with frequency $ \omega_L $ and amplitude $\eta_j$.
\section{\label{sec:Sec3} Heisenberg-Langevin equations}
The HL equations of motion for the optical and mechanical modes in the frame rotating at the laser frequency are, respectively, given by
\begin{align}
{{\dot a}_j} &= \left( {{\rm i}\Delta  - \kappa } \right){a_j} + {\rm i}{g_0}(b_j^\dag  + {b_j}){a_j} + {\rm i}J\left( {{a_{j - 1}} + {a_{j + 1}}} \right)\nonumber \label{Eq:Heisenberg_Equation10}\\
&\qquad\qquad\qquad\qquad\qquad\qquad\qquad\,\,\,+ {\eta _j} - \sqrt \kappa  a_j^{\rm in}(t)\,,\\
{{\dot b}_j} &=  - \left( {{\rm i}{\omega _{\rm m}} + \gamma } \right){b_j} + {\rm i}{g_0}a_j^\dag {a_j} + {\rm i}K({b_{j - 1}} + {b_{j + 1}}) - \sqrt \gamma  b_j^{\rm in}(t)\,,\label{Eq:Heisenberg_Equation20}
\end{align}
where we have defined the laser detuning $ \Delta={\omega _L} - {\omega _c}$.  Besides, $ \kappa $ and $ \gamma $ characterize, respectively, the dissipation of optical and mechanical modes.
The zero-mean value operators $a_{j}^{\rm in}(t)$ and $b_j^{\rm in}(t)$ that describe, respectively, the vacuum optical input noise and the mechanical noise operator, satisfy the commutation relations
\begin{equation}\label{Eq:commutation_relations}
[ { a_{j}^{\rm in}( t), a_{j'}^{{\rm in},\dag }( t')}] = [ {{b_j^{\rm in}}( t),{b_{j'}^{{\rm in},\dag }}( t')}] = \delta_{jj'}\delta (t - t'),
\end{equation}
and the Markovian correlation functions
\begin{align}
\langle {{b_j^{{\rm in},\dag }}( t ){b_{j'}^{{\rm in}}}( t')}\rangle & = {\bar{n}}_{\rm m}\delta_{jj'}\delta (t - t'), \label{Eq:second_order_correlations1}\\
\langle { a_{j}^{{\rm in}}( t) a_{j'}^{{\rm in},\dag }( t')} \rangle  &= \delta_{jj'}\delta ( t - t'),\label{Eq:second_order_correlations}
\end{align}
where we have assumed that each cavity is at zero temperature and $\bar{n}_{\rm m}=[\exp(\hbar \omega_{\rm m} /k_{\rm{B}}T)-1]^{-1}$ is the mean number of thermal phonons of each mechanical mode at temperature $ T $, with $ k_{\rm{B}}$ being the Boltzmann constant.
\subsection{Classical dynamics}
We now employ the mean-field approximation to linearize the dynamics around the classical solutions by decomposing the quantum field operators as $a_j = \alpha_j+ c_j$ and $b_j= \beta_j+ d_j$ where $\alpha_j$ and $ \beta_j$ are the steady-state mean fields describing, respectively, the classical behavior of the optical and mechanical modes, and $c_j$ and $d_j$ are the quantum fluctuations with zero-mean value.
For the aim of this paper, it is enough to consider only the translational symmetry $ \alpha_j=\alpha_{j\pm 1} $ and $ \beta_j=\beta_{j\pm 1} $, which is obtained with an approximately uniform optical driving $\eta_j \simeq \eta$ which therefore excites a background with a small wave vector $ k\approx0 $. Using this assumption, the system dynamics is then simplified to the single-site case. The equations of motion for the steady-state classical mean fields can be obtained by averaging Eqs.~(\ref{Eq:Heisenberg_Equation10}) and (\ref{Eq:Heisenberg_Equation20}) over classical and quantum fluctuations
\begin{align}
{\alpha _j} =\alpha &\simeq  \frac{{{{\rm i} \eta }}}{{\left( {\Delta  + {\rm i}\kappa  + 2J + 2{g_0}  {\mathfrak{R} \beta}} \right)}}\,, \label{Eq:SteadyState10}\\
{\beta _j}=\beta &\simeq \frac{{{g_0}{{\left| {{\alpha }} \right|}^2}}}{{\left( {{\omega _m} - {\rm i}\gamma  - 2K} \right)}}{\mkern 1mu} ,
\label{Eq:SteadyState20}
\end{align}
where $\mathfrak{R} $ denotes the real part.

\subsection{Linearized quantum dynamics}
\begin{figure}
	\begin{center}
		\includegraphics[scale=1]{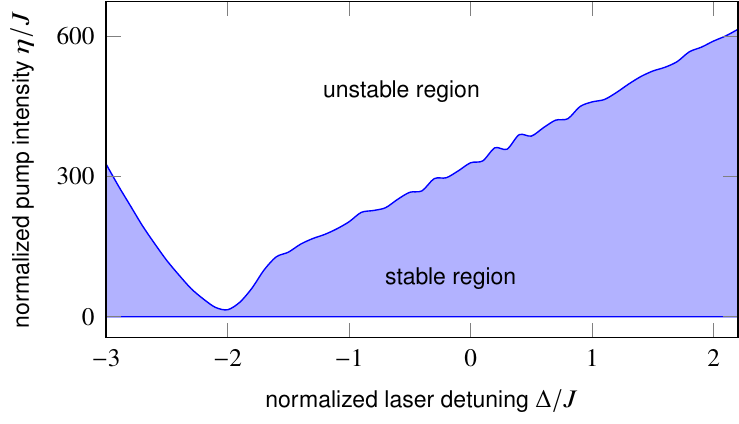}
	\end{center}
	\caption{Stability domain as a function of the normalized input power $ \eta/J $ and normalized detuning $ \Delta/J $. The white and blue areas correspond to the unstable and stable correlated regimes, respectively. The normalized parameters are set with respect to $J$, $\kappa/J=0.1$, $ g_0/J=10^{-4} $, $\gamma/J= 0.002$, $\omega_{\rm m}/J=0.1$ and $K/J=0.05$. Temperatures of the photonic and phononic heat baths are considered to be zero.}
	\label{Fig:Fig2}
\end{figure}
We study the quantum statistical properties of the system through the small fluctuations of the operators around the steady-state classical mean values given by Eqs.~(\ref{Eq:SteadyState10}) and (\ref{Eq:SteadyState20}). Using the standard definition of the optical and mechanical mode quadratures $ {{ X}_j} = ({{{ c}_j} +  c_{j}^\dag })/{{\sqrt 2 }}$, ${{ Y}_j} = ({{{ c}_j} -  c_{j}^\dag })/{{{\rm i}\sqrt 2 }}$, ${{ x}_j} = ({{{ d}_j} +  d_{j}^\dag })/{{\sqrt 2 }}$ and ${{ y}_j} = ({{{ d}_j} -  d_{j}^\dag })/{{{\rm i}\sqrt 2 }}$, the equations of motion for the quantum fluctuations are given by
\begin{align}\label{key}
{{\dot X}_j} &=  - \left( {\Delta  + 2 {g_0} \mathfrak{R}{\beta}} \right){Y_j}  - \kappa {X_j} - 2{g_0}\mathfrak{I}{\alpha}{x_j} \nonumber\\& \qquad \qquad \qquad \quad\quad - J\left( {{Y_{j - 1}} + {Y_{j + 1}}} \right) - \sqrt \kappa  X_j^{{\rm{in}}}\left( t \right){\mkern 1mu} ,\\
{{\dot Y}_j} &= \left( {\Delta  + 2 {g_0}\mathfrak{R} {\beta}} \right){X_j} - \kappa {Y_j} + 2{g_0}\mathfrak{R} {\alpha}{x_j} \nonumber\\& \qquad \qquad \qquad \qquad\quad+ J\left( {{X_{j - 1}} + {X_{j + 1}}} \right) - \sqrt \kappa  Y_j^{{\rm{in}}}\,,\\
{{\dot x}_j} &=  - \gamma {x_j} + {\omega _m}{y_j} - K\left( {{y_{j - 1}} + {y_{j + 1}}} \right) - \sqrt \gamma  x_j^{{\rm{in}}}{\mkern 1mu} ,\\
{{\dot y}_j} &=  - {\omega _m}{x_j} - \gamma {y_j} + 2{g_0}\left( {\mathfrak{R} \alpha {X_j} + \mathfrak{I} \alpha {Y_j}} \right) \nonumber\\& \qquad \qquad \qquad \quad\qquad + K\left( {{x_{j - 1}} + {x_{j + 1}}} \right) - \sqrt \gamma  y_j^{{\rm{in}}}{\mkern 1mu} ,
\end{align}
where $\mathfrak{I} $ denotes the imaginary part. We now express the linearized HL equations in the following compact matrix form
\begin{equation}\label{eq:LinearizedLangevin}
\dot {\bf u}(t)={\bf A}{\bf u}(t)+{\bf n}(t)\,,
\end{equation}
where we have defined the vector of fluctuation operators $ {\bf u} = {\left[ \cdots	{{{\bf v}}_{j - 1}},{{{ {\bf v}}_j}},	{{\bf v}_{j + 1}},\cdots \right]^T}$ with $ {\bf v}_j = {[{X}_j,{Y}_{j},x_j,y_j]}$ and the corresponding vector of noises $ {\bf n} = {\left[ \cdots	{{{{\bf m}}_{j - 1}}},{{{ {\bf m}}_j}},	{{{{\bf m}}_{j + 1}}},\cdots \right]^T}$ with $ {\bf m}_j = {[\sqrt \kappa X_j^{\rm in},\sqrt \kappa Y_{j}^{\rm in},\sqrt \gamma x_j^{\rm in},\sqrt \gamma y_{j}^{\rm in} ]}$, in which  $ {{ X}_j^{\rm in}} = ({{{ a}_j^{\rm in}} +  a_{j}^{\dag,{\rm in}} })/{{\sqrt 2 }}$, ${{ Y}_j^{\rm in}} = ({{{ a}_j^{\rm in}} -  a_{j}^{\dag,{\rm in}} })/{{{\rm i}\sqrt 2 }}$, ${{ x}_j^{\rm in}} = ({{{ b}_j^{\rm in}} +  b_{j}^{\dag,{\rm in}} })/{{\sqrt 2 }}$ and ${{ y}_j^{\rm in}} = ({{{ b}_j^{\rm in}} -  b_{j}^{\dag,{\rm in}} })/{{{\rm i}\sqrt 2 }}$ are the input noise quadratures of the optical and mechanical modes. Furthermore, we define the drift matrix ${\bf A}$ as
\begin{equation}\label{key}
{\bf A} = \left[ {\begin{array}{*{20}{c}}
	\ddots & \ddots &0&0&0\\
	\ddots &{{{\bf B}}}&{\bf C}&0&0\\
	0&{\bf C}&{{{\bf B}}}&{\bf C}&0\\
	0&0&{\bf C}&{{{\bf B} }}& \ddots \\
	0&0&0& \ddots & \ddots
	\end{array}} \right]\,,
\end{equation}
with the blocks
\begin{equation}\label{key}
{\bf B} = \left[ {\begin{array}{*{20}{c}}
	{ - \kappa }&-{\left( {\Delta  +2 {g_0}\mathfrak{R}  {\beta }} \right)}&{ - 2{g_0}\mathfrak{I} {\alpha }}&0\\
	{  \left( {\Delta  +2 {g_0}\mathfrak{R} {\beta }} \right)}&{ - \kappa }&{2{g_0}\mathfrak{R} {\alpha }}&0\\
	0&0&{ - \gamma }&{{\omega _m}}\\
	{2{g_0}\mathfrak{R} \alpha }&{2{g_0}\mathfrak{I} \alpha }&{ - {\omega _m}}&{ - \gamma }
	\end{array}} \right]\,,
\end{equation}
\begin{equation}\label{key}
{\bf C} = \left[ {\begin{array}{*{20}{c}}
	0&-J&0&0\\
	{  J}&0&0&0\\
	0&0&0&-K\\
	0&0&{  K}&0
	\end{array}} \right]\,.
\end{equation}
\section{\label{sec:Sec4} Steady-state quantum correlations}
Due to the Gaussian nature of the quantum noises and to the linearized dynamics, the steady state of the quantum fluctuations of the OMCs is a continuous variable $ 2N $-partite Gaussian state, which is completely determined by its $ 4 N\times 4N $ CM. The formal solution of Eq.~(\ref{eq:LinearizedLangevin}) is
\begin{equation}\label{key}
{\bf u}(t) = {\bf M}(t)  {\bf u}(0) + \int\limits_0^t {{{\bf M}(t-s)}{{ {\bf n}}}(s)ds} \,,
\end{equation}
with ${\bf M}(t)= \exp[{{t{\bf A}}}] $. The CM defined as
\begin{equation}
{{\bf V}_{pq}}(t) = \frac{1}{2}\left\langle {{{ {\bf u}}_p}(t){{ {\bf u}}_q}(t) + {{ {\bf u}}_q}(t){{ {\bf u}}_p}(t)} \right\rangle \, ,
\end{equation}
contains all information about the quantum correlation between various mechanical and optical modes where $ { {\bf u}}_p(t) $ is the $p$th component of the vector ${{\bf u}}(t)$.

The system reaches its steady state when ${\bf M}(\infty)=0$. Our analysis is restricted to the stable regime where all the eigenvalues of the drift matrix have negative real parts. In Fig.~\ref{Fig:Fig2}, we plot the region of stability as a function of the normalized laser pump intensity and detuning. For large laser drive, the system enters the unstable region. In the steady state, one gets the CM elements as
\begin{equation}\label{key}
{\bf V}_{ij} = \sum\limits_{k,l} \int\limits_0^\infty  {ds\int\limits_0^\infty  {ds'{{\bf M}_{ik}}(s){{\bf M}_{jl}}( s')\Phi _{kl}(s - s')}},
\end{equation}
where
\begin{align}
\Phi _{kl}(s) &=\frac{1}{2}\! \left\langle {{ {\bf n}_k}(s) {\bf n}_l(s') +  {\bf n}_l(s'){ {\bf n}_k}(s)} \right\rangle \nonumber\\
&= {\bf D}_{kl}\delta ( s - s'),
\end{align}
where ${\bf D} = {\rm diag} {\left[ \cdots	{\bf F},{\bf F},{\bf F},\cdots \right]^T}$ with $ {\bf F}= {\rm diag} {[\kappa,\kappa,\gamma(2 \bar n_{\rm m}+1),\gamma(2 \bar n_{\rm m}+1)]}$. When the stability conditions are satisfied so that $ {\bf M}(\infty)=0$, the steady-state CM, $ {\bf V} $, can be obtained by solving the linearized HL equation~(\ref{eq:LinearizedLangevin}) for the quantum fluctuations, which fullfil the following Lyapunov equation
\begin{equation}
{{\bf A}}{\bf V} + {\bf V}{{\bf A}}^T = - {\bf D}\,.
\end{equation}
With these classical and quantum steady-state solutions in hand, we next employ the CM formalism to calculate the steady-state quantum correlations. We check the presence of the quantum correlations between the mechanical and optical modes on the same site, as well as between the mechanical or optical modes with different site indices. Considering the following reduced CM of the two modes
\begin{equation}
{\bf V}_R = \left[ {\begin{array}{*{20}{c}}
	{{{\bf V}_{A}}}&{{{\bf V}_{C}}}\\
	{{{\bf V}_{C}^T}}&{{{\bf V}_{B}}}
	\end{array}} \right]\,,
\end{equation}
one can calculate the quantum correlations. Here, ${{\bf V}_{A}}$, ${{\bf V}_{B}}$ and ${{\bf V}_{C}}$ are $2 \times 2$  matrices where ${{\bf V}_{A}}$ and ${{\bf V}_{B}}$  account for the local properties of modes  $A$ and $B$, respectively, while  ${\bf V}_{C}$  describes intermode correlations. $ A $ and $ B $ may stand for two different modes.
\begin{figure}
	\includegraphics[width=9cm]{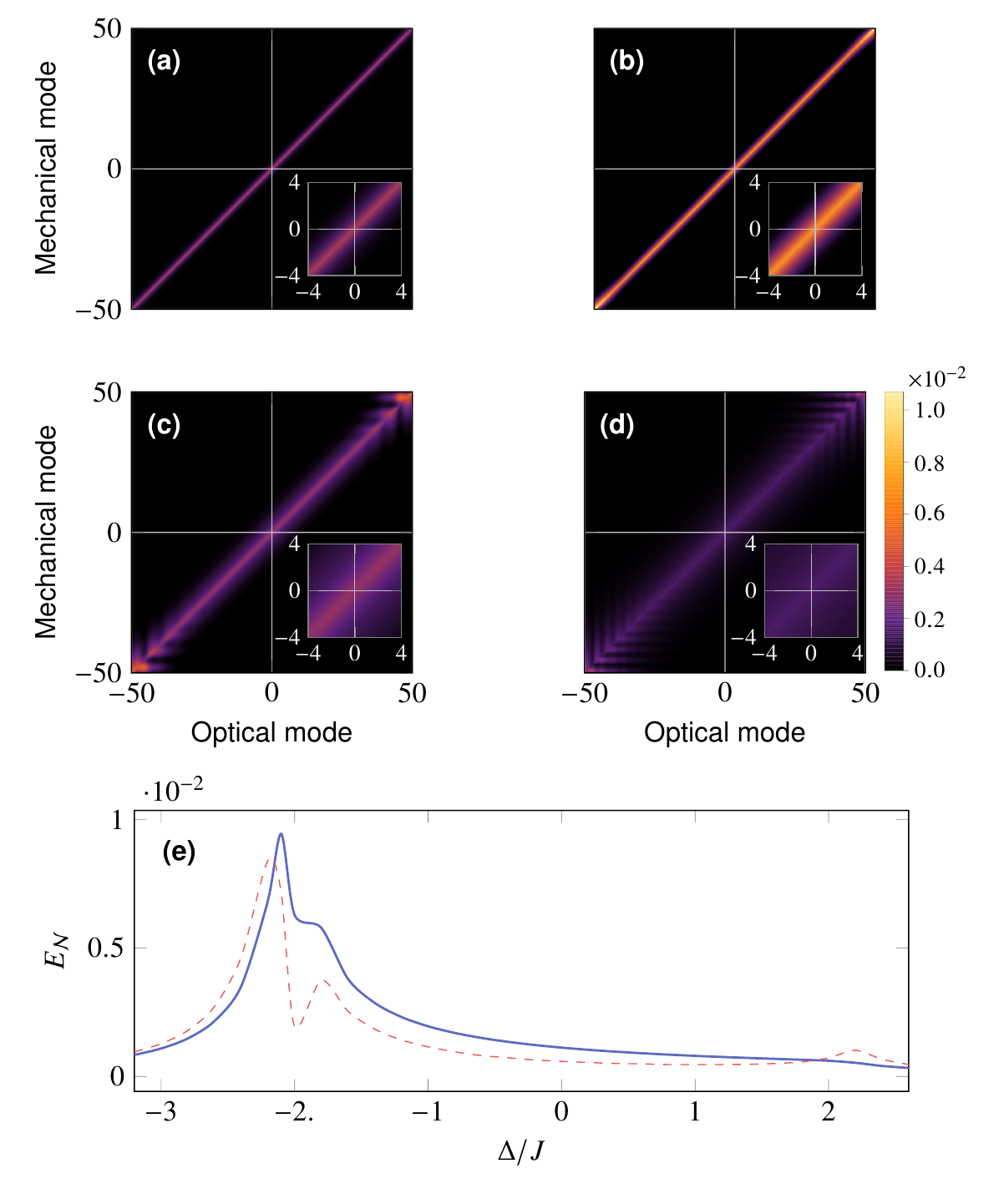}
	\caption{The degree of entanglement between optical and mechanical modes in terms of the logarithmic negativity for various values of the laser detuning: (a) $ \Delta/J=-2.5 $, (b) $ \Delta/J=-2.1 $, (c) $ \Delta/J=-1.7 $ and (d) $ \Delta/J=-1.3 $ for 101 coupled OMSs. (e) The logarithmic negativity between the two optical and mechanical modes with the same site index $ j=-50 $ or $ j=50 $  (blue solid line) and $ j=0 $ (red dashed line) versus the laser detuning. We set normalized parameters with respect to $J$, $\kappa/J=0.1$, $\eta/J=15$, $ g_0/J=10^{-4} $, $\gamma/J= 0.002$, $\omega_{\rm m}/J=0.1$ and $K/J=0.05$. Temperatures of the photonic and phononic heat baths are considered to be zero.}
	\label{Fig:Fig3}
\end{figure}
\subsection{Steady-state entanglement}
We quantify the degree of entanglement in terms of the logarithmic negativity, which is an entanglement monotone, and it is given by ${E_N} = \max \{ 0, - \ln 2{\tilde \nu_- }\} $  with   $ \tilde{\nu}_- = 2^{-1/2}  \left( \Sigma _-  - \sqrt {\Sigma _ - ^2 - 4\det {\bf V}_R}\right)^{1/2} $ being the smallest of the two symplectic eigenvalues of the partially transposed transposed CM and  ${\Sigma _ \pm } = \det {\bf V}_{A} + \det {\bf V}_{B} \pm 2\det {\bf V}_{C}$.


\subsubsection{Photon-phonon entanglement}
 The degree of entanglement between optical and mechanical modes in terms of the logarithmic negativity for various laser detuning at zero temperature of both the photonic and phononic heat baths is shown in Fig.~\ref{Fig:Fig3}~(a)-(d). We can see that one has mostly on-site optomechanical entanglement, i.e., between modes at the same sites, and that there is no long-range photon-phonon entanglement. However, as suggested by the zoomed insets, one has that, due to the combined action of the on-site optomechanical interaction and of tunneling of the photons and phonons between lattice sites, there is some amount of off-site entanglement between optical and mechanical modes. For instance, the optical mode at the site $ j=0 $ is entangled with the neighbor mechanical modes at sites $|j|<1$, $|j|< $, $|j|<3$, and $|j|<4$ for $ \Delta/J=-2.5 $, $ \Delta/J=-2.1 $, $ \Delta/J=-1.7 $ and  $ \Delta/J=-1.3 $. It is also evident the detuning has a significant effect on the optomechanical entanglement. We address this issue in Fig.~\ref{Fig:Fig3}~(e) where we have plotted the logarithmic negativity between the two optical and mechanical modes with the same site index $ j=-50 $ or $ j=50 $ and $ j=0 $ versus the laser detuning.
 Our choices for the detuning and laser-drive intensity correspond to the stable region of Fig.~\ref{Fig:Fig2}. Furthermore, since we did not consider the periodic boundary conditions one can see a non-uniform behavior at the lattice edges.

\begin{figure}
	\includegraphics[width=9cm]{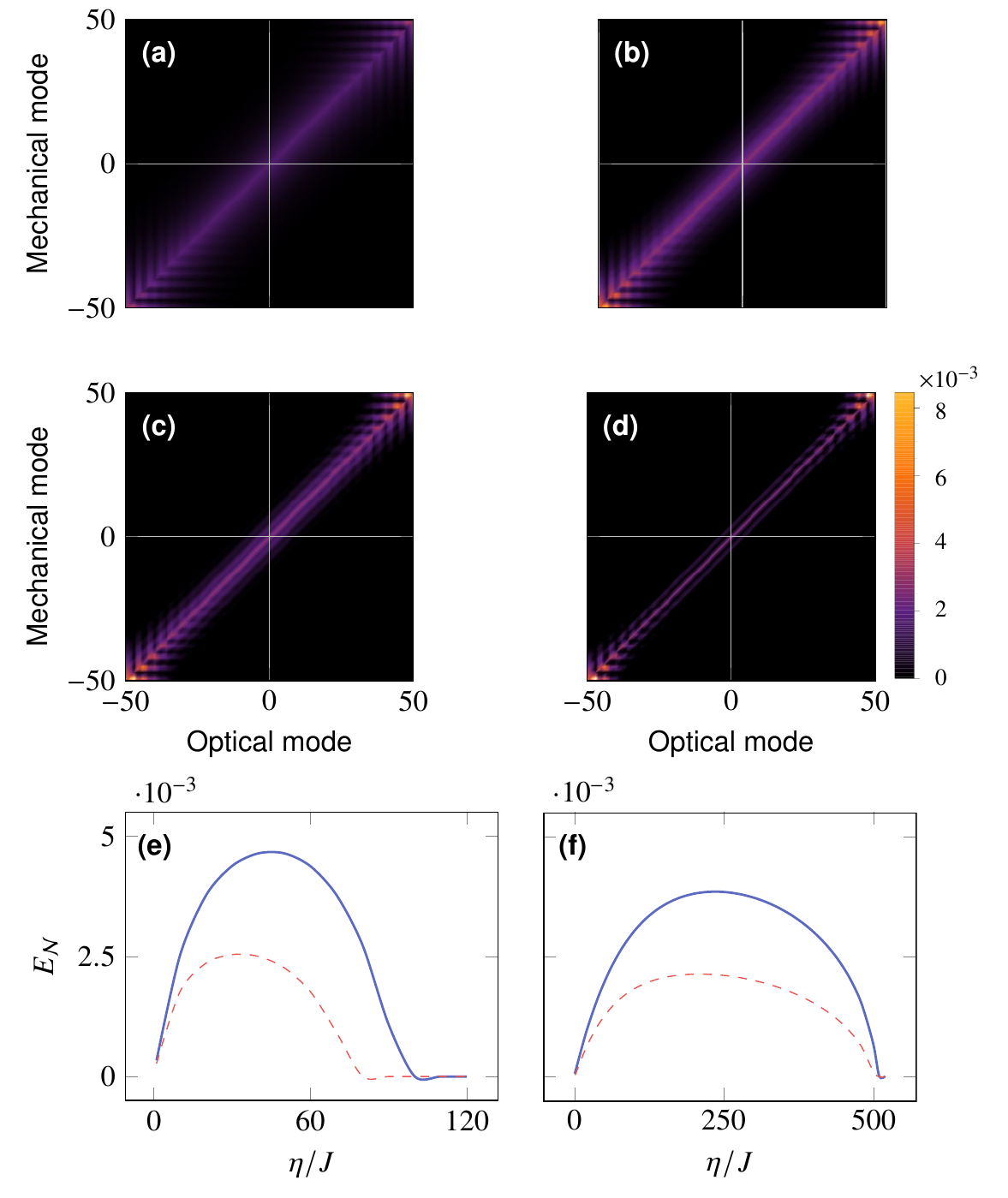}
	\caption{The degree of entanglement between optical and mechanical modes in terms of the logarithmic negativity for various values of the laser intensity: (a) $ \eta/J=50 $, (b) $ \eta/J=150 $, (c) $ \eta/J=250 $ and (d) $ \eta/J=350 $ for 101 coupled OMSs. We set $\Delta/J=1.5$, other parameters are the same as Fig.~\ref{Fig:Fig3}. Panels (e) and (f) show the logarithmic negativity between the two optical and mechanical modes with the same index $ j=-50 $ or $ j=50 $ (blue solid line) and $ j=0 $ (red dashed line)  versus the laser-drive intensity
		for two values of the laser detuning: (e) $ \Delta/J=-1.5 $ and  (f) $ \Delta/J=1.5 $.}
	\label{Fig:Fig4}
\end{figure}
In Fig.~\ref{Fig:Fig4}~(a)-(d), we show how the photon-phonon entanglement varies as a function of the laser pump intensity for a fixed laser detuning, $\Delta/J=1.5$. By increasing the laser intensity the entanglement first tends to increase and then to decrease as we approach the unstable region. Therefore,  there is a non-monotonic behavior of on-site entanglement. We show this fact in Figs~\ref{Fig:Fig4}(e) and \ref{Fig:Fig4}(f) where we have plotted the logarithmic negativity between the two optical and mechanical modes with the same site index at the lattice edge ($ j=-50$ or $50 $) and at the lattice center ($ j=0 $) versus the laser-drive intensity for two values of the laser detuning.

Finally we have also studied the eventual presence of photon-photon or phonon-phonon entanglement between different sites. We have verified that for all choices of the parameters this kind of inter-site entanglememt is always zero.

\begin{figure}
	\includegraphics[scale=1]{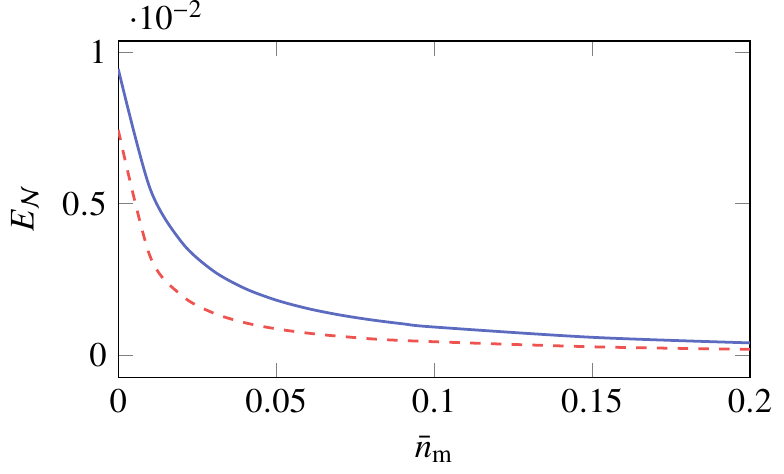}
	\caption{
		Steady-state photon-phonon entanglement for the site index $ j=-50 $ or $ j=50 $ versus the thermal phonon number $\bar{n}_{\rm m}$ for two values of the normalized laser detuning $ \Delta/J=-2 $ (red dashed line) and $ \Delta/J=-2.1 $ (blue solid line) for 101 coupled OMSs. We have considered here mechanical resonators with frequency $\omega_m/2\pi = 9~\rm{GHz}$ and the other parameters are the same as in Fig.~\ref{Fig:Fig3}. }
	\label{Fig:Fig5}
\end{figure}
\subsubsection{Thermal effects on the generated entanglement}
Usually, quantum correlations and entanglement in particular are fragile with respect to thermal noise. Therefore, the investigation of the effect of thermal fluctuations on the bipartite quantum correlations in OMCs is of particular relevance for applications.

In Fig.~\ref{Fig:Fig5}, we show how the on-site photon-phonon entanglement changes with increasing thermal phonon number $\bar{n}_{\rm m}$. Evidently, the on-site photon-phonon entanglement decays for increasing temperatures and it persists at ultra-cryogenic temperatures achievable in dilution refrigerators (for example $ \bar n_{\rm m}\simeq 0.06 $ for mechanical resonance frequencies $\omega_m/2\pi = 9~\rm{GHz}$  at a temperature of $ T = 0.15 K $).

\begin{figure}
	\includegraphics[width=9cm]{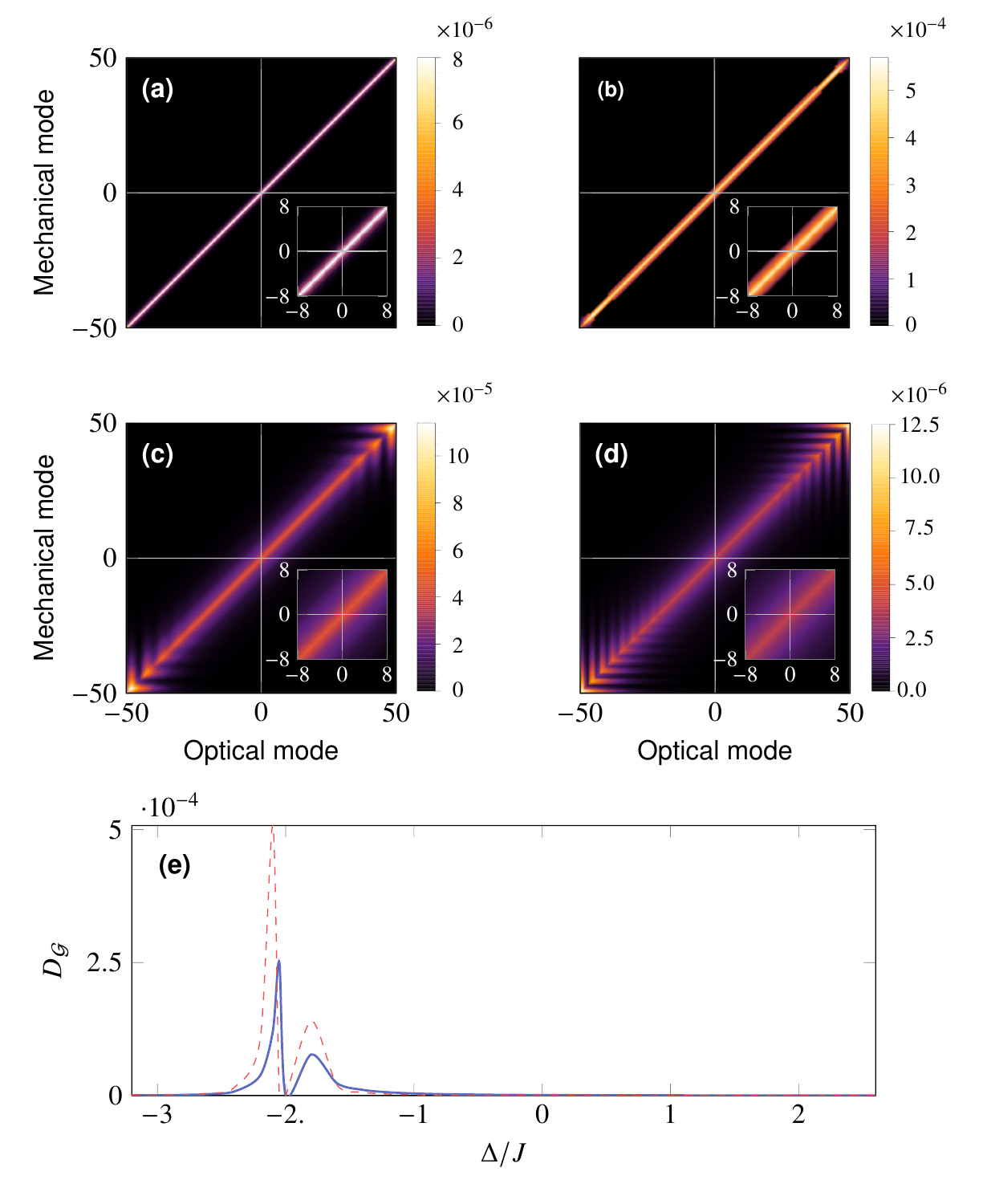}
	\caption{Steady-state symmetrized Gaussian quantum discord between optical and mechanical modes for various laser detuning values: (a) $ \Delta/J=-2.5 $, (b) $ \Delta/J=-2.1 $, (c) $ \Delta/J=-1.7 $ and (d) $ \Delta/J=-1.3 $ for 101 coupled OMSs. Panel (e) shows the symmetrized Gaussian quantum discord between the two optical and mechanical modes with the same index $ j=0 $ (blue solid line) and $ j=50 $ (red dashed line) versus the laser detuning. The heat bath temperatures for mechanical and optical modes are considered to be zero. Other parameters are the same as Fig.~\ref{Fig:Fig3}.}
	\label{Fig:Fig6}
\end{figure}
\begin{figure}[t]
	\includegraphics[width=9cm]{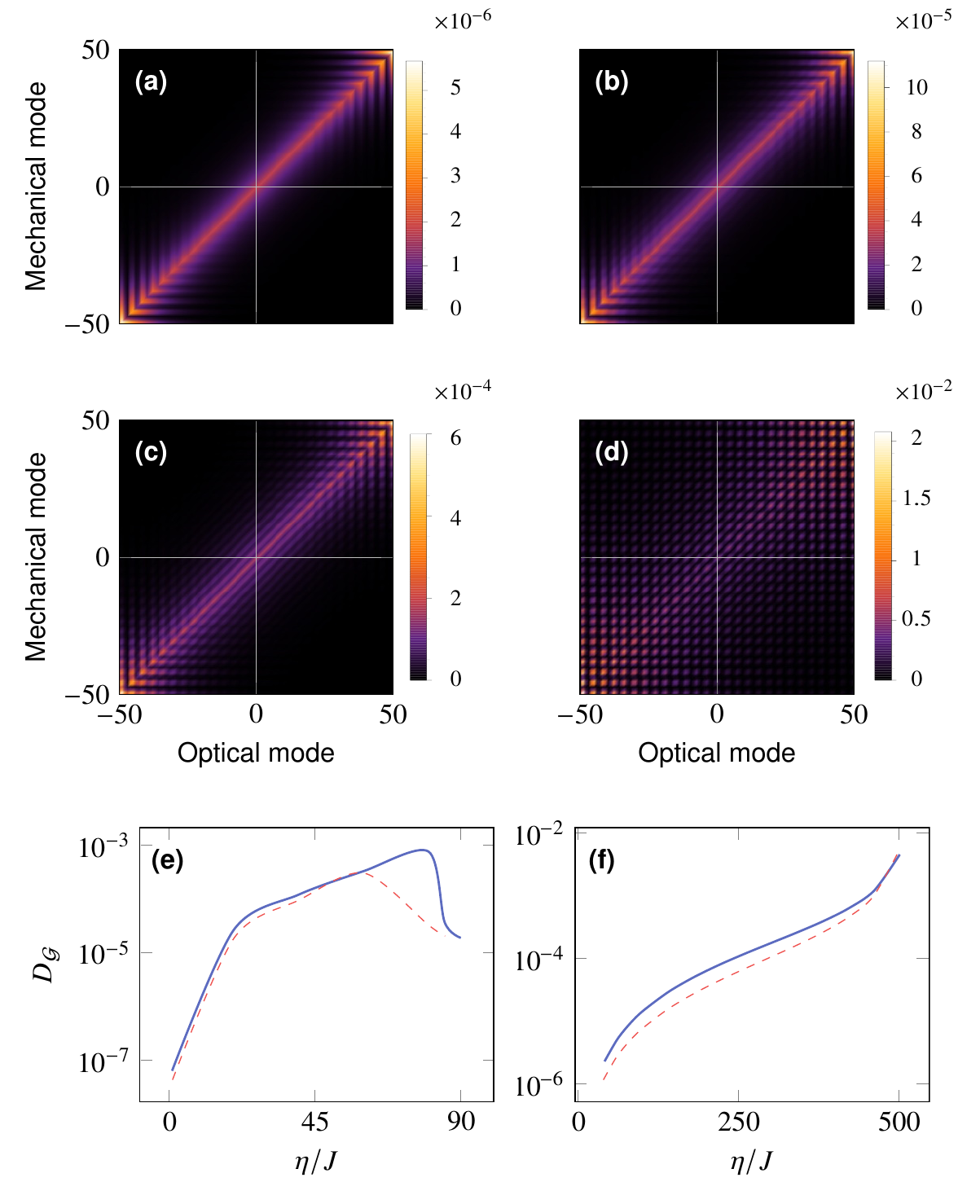}
	\caption{Steady-state Gaussian quantum discord between optical and mechanical modes for various laser intensity values: (a) $ \eta/J=50 $, (b) $ \eta/J=200 $, (c) $ \eta/J=350 $ and (d) $ \eta/J=500 $ for 101 coupled OMSs. Here, we set $\Delta/J=1.5$, and other parameters are the same as Fig.~\ref{Fig:Fig3}. Panels (e) and (f) show the Steady-state Gaussian quantum discord  between the two optical and mechanical modes with the same site index $ j=50 $ or $ j=-50 $ (red dashed line) and $ j=0 $ (blue solid line)  versus the laser-drive intensity
		for two values of the laser detuning: (e) $ \Delta/J=-1.5 $ and  (f) $ \Delta/J=1.5 $.}
	\label{Fig:Fig7}
\end{figure}
\subsection{Steady-state Gaussian quantum discord}
\begin{figure}
	\includegraphics[width=9cm]{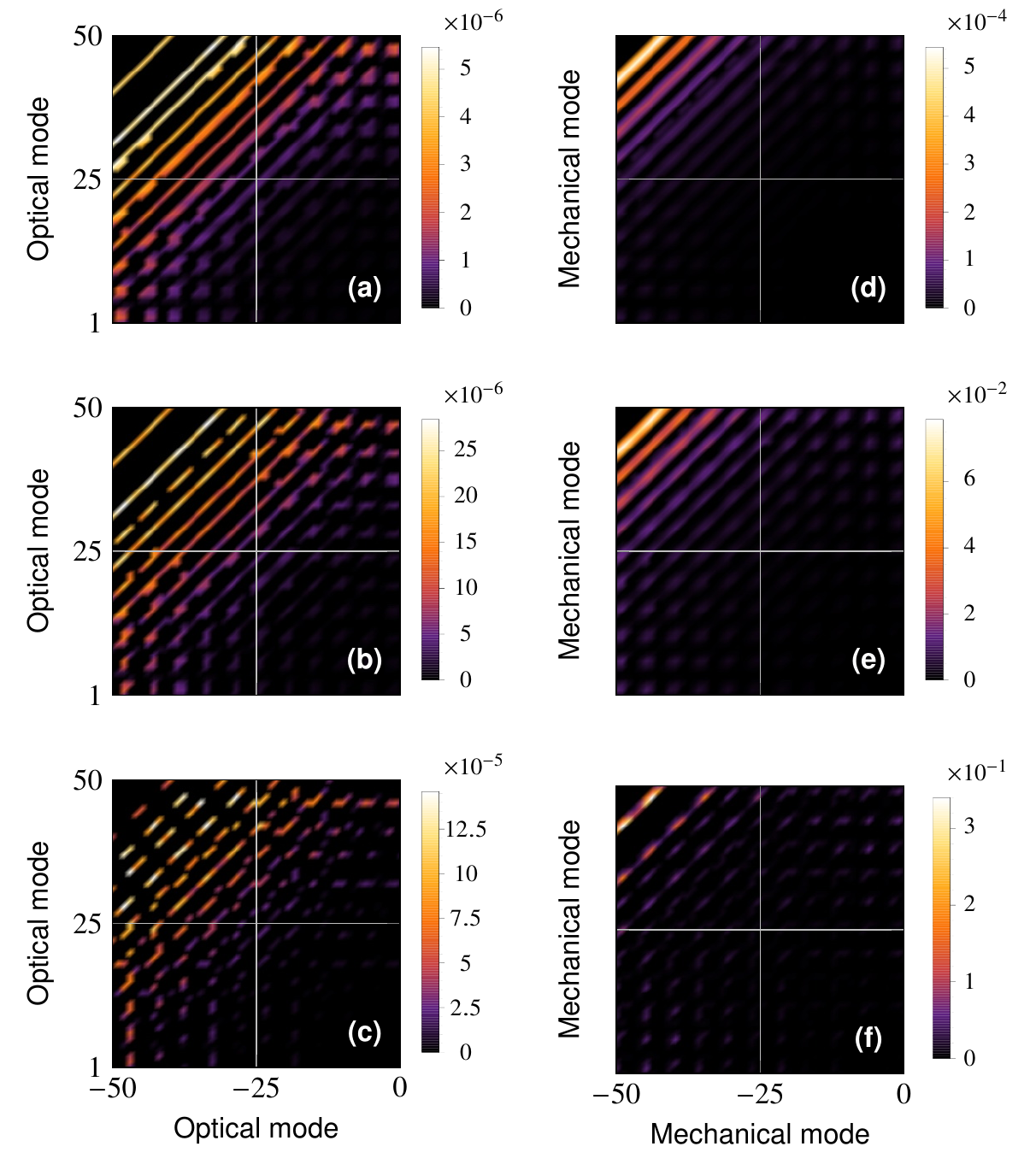}
	\caption{Steady-state Gaussian quantum discord between (a)-(c) optical modes and (e)-(f) mechanical modes for various values of the laser intensity: (a) and (d) $ \eta/J=80 $, (b) and (e) $ \eta/J=100 $, and (c) and (f) $ \eta/J=120 $ for 101 coupled OMSs. The normalized laser detuning is set $ \Delta/J=-1.5 $. Other parameters are the same as Fig.~\ref{Fig:Fig2}.
	}
	\label{Fig:Fig8}
\end{figure}
\begin{figure}
	\includegraphics[width=9cm]{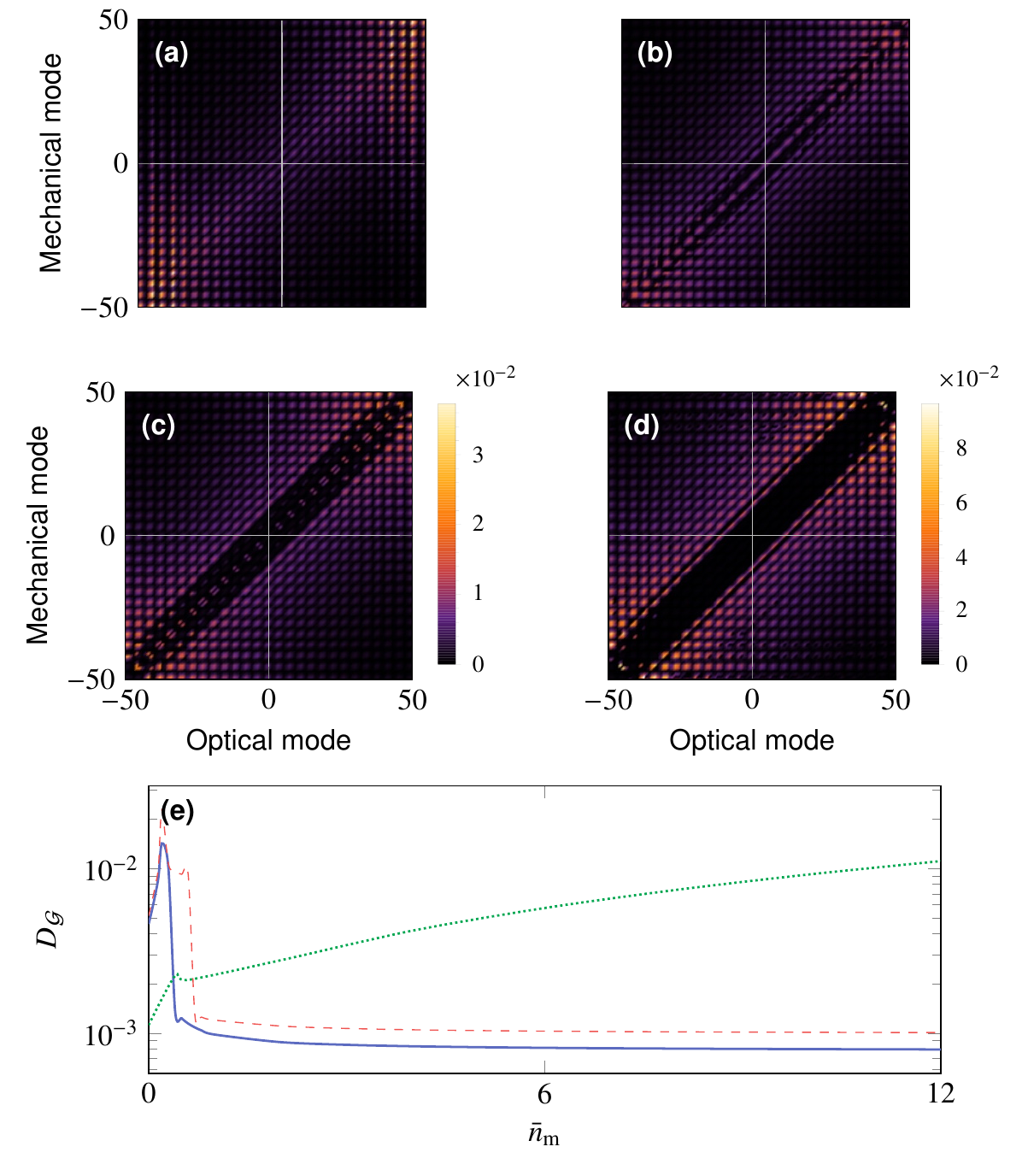}
	\caption{
		Steady-state Gaussian quantum discord under different heat bath phonon number for normalized laser detuning $ \Delta/J=1.5 $, driving $ \eta/J=500 $, and mechanical resonance frequency $\omega_m/2\pi = 9~\rm{GHz}$:  (a) $ \bar n_{\rm m}=0.1 $ ($ T=0.18 K $), (b) $ \bar n_{\rm m}=0.5 $ ($ T=0.39 K $), (c) $ \bar n_{\rm m}=2.5 $ ($ T=1.28 K $), and (d) $ \bar n_{\rm m}=12.5 $ ($ T=5.59 K $) for 101 coupled OMSs. Legend bar is the same for (a)-(c).
Parameters are the same as Fig.~\ref{Fig:Fig2}. See Fig.~\ref{Fig:Fig7}(d) for zero heat-bath temperature. Panel (e) shows the Steady-state Gaussian quantum discord  between the two optical and mechanical modes with the same site index $ j=50 $ or $ j=-50 $ (red dashed line), $ j=0 $ (blue solid line) and with the different site index $ j=0 $ and $ j=50 $ (green dotted line) versus heat bath phonon number. }

		\label{Fig:Fig9}
\end{figure}
It is also interesting to examine if quantum discord \cite{Zurek,Henderson}, a measure of the quantumness of correlations, is present in the steady state of the system. The Gaussian quantum discord is an asymmetric quantity and the Gaussian quantum A-discord of the Gaussian state of two modes, $A$ and $B$, is given by \cite{Adesso,Giorda}
\begin{equation}
{\mathfrak{D}}^\rightarrow = f\left( {\sqrt {\beta} } \right) - f\left( {{\upsilon _ - }} \right) - f\left( {{\upsilon _ + }} \right) - f\left( {\sqrt \varepsilon  } \right)
\end{equation}
where
\begin{equation}
f\left( x \right) \!=\! \left( {\frac{{x \!+\! 1\!}}{2}} \right)\log_{10} \left( {\frac{{x \!+\! 1\!}}{2}} \right) \!-\! \left( {\frac{{x \!-\! 1\!}}{2}} \right) \log_{10} \left( {\frac{{x \!-\! 1\!}}{2}} \right),
\end{equation}
\begin{equation}
{\upsilon _ \pm } =\sqrt{\dfrac{{\Sigma _+ } \pm \sqrt {{\Sigma_+^{ 2}} - 4\det {\bf V}_R}}{2}  }
\end{equation}
are the two symplectic eigenvalues of the two-mode CM and
	\begin{equation}
\varepsilon  = \left\{ {\begin{array}{*{20}{c}}
	{\frac{{2{\gamma ^2} + \left( {\beta  - 1} \right)\left( {\delta  - \alpha } \right) + 2\left| \gamma  \right|\sqrt {{\gamma ^2} + \left( {\beta  - 1} \right)\left( {\delta  - \alpha } \right)} }}{{{{\left( {\beta  - 1} \right)}^2}}},}&{\frac{{{{\left( {\delta  - \alpha \beta } \right)}^2}}}{{\left( {\beta  + 1} \right){\gamma ^2}\left( {\alpha  + \delta } \right)}} \le 1;}\\
	{\frac{{\alpha \beta  - {\gamma ^2} + \delta  - \sqrt {{\gamma ^4} + {{\left( {\delta  - \alpha \beta } \right)}^2} - 2{\gamma ^2}\left( {\delta  + \alpha \beta } \right)} }}{{2\beta }},}&{{\rm{otherwise}},}
	\end{array}} \right.
\end{equation}
where  $\alpha  = {{\mathop{\rm det {\bf V}}\nolimits} _A}$, $\beta  = {{\mathop{\rm det {\bf V}}\nolimits} _B}$, $\gamma  = {{\mathop{\rm det {\bf V}}\nolimits} _C}$ and $\delta  = \det {\bf V}_R$ are  the symplectic invariants. One can obtain the Gaussian quantum B-discord $ {\mathfrak{D}}^\leftarrow $ by swapping the roles of the two modes, $A$ and $B$, which is equivalent to swap $\alpha$ and $\beta$ in the above formulas. Since we are interested in quantum correlations in general between the different modes in the one-dimensional array, from now on we will consider the symmetrized quantum discord, $D_{\mathcal{G}}= {\rm max} \left\lbrace   {\mathfrak{D}}^\leftarrow , {\mathfrak{D}}^\rightarrow  \right\rbrace   $.
\subsubsection{Photon-phonon steady-state Gaussian quantum discord}
Fig.~\ref{Fig:Fig6} shows the behavior of the symmetrized quantum discord $D_{\mathcal{G}}$ for various laser detuning values at zero temperature of both photonic and phononic modes. Similarly to what occurred for entanglement, changing the laser detuning has a significant effect on the photon-phonon Gaussian quantum discord, and again we have a similar behavior with that of entanglement with the above choice of parameters, with the presence of larger on-site discord between the mechanical and the optical mode and which extends for few sites. One starts to see a different behavior between Gaussian discord and entanglement when looking at the dependence upon the driving power and specifically if we consider increasing values of the laser drive $\eta $.
In Fig.~\ref{Fig:Fig7}, we show how steady-state photon-phonon Gaussian quantum discord varies with the laser intensity for a fixed laser detuning, $\Delta/J=1.5$. In contrast with the behavior of entanglement, we have that by increasing the laser intensity one has a significant increase of Gaussian quantum discord between optical and mechanical sites (see Figs.~\ref{Fig:Fig7}(e) and \ref{Fig:Fig7}(f)). Moreover, at larger values one can see a long-range correlation between optical and mechanical modes appearing (see Fig.~\ref{Fig:Fig7}(d)).

\subsubsection{Photon-photon and phonon-phonon steady-state Gaussian quantum discord}
The appearance of long-range quantum correlations occurs also when considering either only optical modes or only mechanical modes, at each site of the OMC, in clear contrast with the case of entanglement which is instead completely absent, even between neighboring sites. This fact is shown in Fig.~\ref{Fig:Fig8}. As can bee seen, for a fixed laser detuning, by increasing the laser intensity the steady-state Gaussian quantum discord between modes of the same nature increases.

\subsubsection{Thermal effects on the steady state Gaussian quantum discord}
It is relevant to study the robustness of the Gaussian quantum discord with respect to temperature as we did it already for entanglement. The steady-state Gaussian quantum discord under different heat-bath phonon number for normalized laser detuning $ \Delta/J=1.5 $ and laser intensity $ \eta/J=500 $ is depicted in Fig.~\ref{Fig:Fig9}. One can see a non-monotonic behavior in Gaussian quantum discord by increasing the thermal phonon number. It first tends to increase, then decreases and finally increases again. This behavior is somehow unexpected and it can be regarded as the evidence of thermally induced Gaussian quantum discord in OMCs. This is not completely novel however in quantum many-body systems; for instance, the transverse-field $ XY $ model, also shows non-monotonic behavior of its quantum correlations (for instance see \cite{Chanda2018} and references therein). We remark however that our model is not exactly the same as $ XY $ model for what concerns the effects of the thermal environment because in the latter the involved excitations has similar frequencies and therefore similar thermal effects, while in our case, due to the large difference in frequencies between optical and mechanical modes, only the phonon modes are appreciably affected by a nonzero reservoir temperature. The phenomenon investigated here shares instead some similarity with what has been already underlined in \cite{Ciccarello,Korolkova} where it has been shown that for continuous-variable bipartite systems, quantum discord can increase for increasing thermal noise because they represent nonclassical correlations which are induced and maintained thanks to the mediating action of the local dissipative bath.

\section{\label{sec:Sec5} Conclusions}
In conclusion, our investigation clearly demonstrates the presence of appreciable quantum correlations in an OMC where each site consists of two localized, optical and mechanical, modes coupled locally via the optomechanical interaction. The modes of nearby sites are connected via both photon and phonon tunneling. In particular, the generation of on-site or short-range entanglement between optical and mechanical modes that rely on the optomechanical interactions in OMCs seems feasible at ultracryogenic temperatures. The generated entanglement is very fragile with respect to thermal noise. We have also shown that there is no long-range entanglement between optical and mechanical modes. Moreover, there is no photon-photon or phonon-phonon entanglement in the system. 
For what concerns the absence of strong entanglement between modes of the same nature, this is due to the quantum dynamics realized by the chosen model Hamiltonian). In fact, it does not contain terms of the form of $ a_j^\dagger a_{j\pm1}^\dagger+a_j a_{j\pm1} $ for the photonic modes (or $ b_j^\dagger b_{j\pm1}^\dagger+b_j b_{j\pm1} $ for phononic modes). It only contains hopping terms which cannot directly entangle modes of the same nature.

We have then examined a weaker form of quantum correlation, i..e., Gaussian quantum discord, and we have studied if quantum discord is present in the steady-state of the system for various control parameters. The Gaussian quantum discord behavior is completely different, one has long-range features in all the three possible cases of correlations, i.e., photon-phonon, photon-photon, and phonon-phonon, at variance with what occurs with entanglement. A further interesting aspect is the thermal activation of quantum discord, i.e., the fact that photon-phonon discord \emph{increases} with increasing temperature. In our opinion this is a manifestation of the transfer of nonclassical correlations mediated by the thermal reservoir, as already discussed for continuous variable systems in \cite{Ciccarello,Korolkova}.

The present study which paves the way toward the investigation of many-body entanglement, can be considered as the first step toward controlled quantum correlations between different quantum processors across the lattice sites with potential applications in quantum information possessing and storage. The proposed scheme also provides a suitable platform for quantum simulation of various many-body systems with optomechanical crystals by tuning the system parameters.

It should be noted that we did not consider the disorder effect in our study. As an outlook, the system under consideration can be generalized to a more realistic case where the lattice disorder is also present in the system. Another outlook may be the generalization to the case of two-dimensional lattices of coupled optomechanical systems.

\begin{acknowledgments}
We would like to thank the Vice President for Research of the University of Isfahan for its support. DV acknowledges the support of the European Union Horizon 2020 Programme for Research and Innovation through the Project No. 732894 (FET Proactive HOT).
\end{acknowledgments}

\end{document}